# PERFORMANCE CHARACTERISTICS OF POSITIVE AND NEGATIVE DELAYED FEEDBACK ON CHAOTIC DYNAMICS OF DIRECTLY MODULATED InGaAsP SEMICONDUCTOR LASERS

Bindu M. Krishna[1], Manu. P. John[*], V. M. Nandakumaran[*]
Sophisticated Test and Instrumentation Centre
[*]International School of Photonics
Cochin University of Science and Technology, Cochin-682022, Kerala, India

**ABSTRACT**
The chaotic dynamics of directly modulated semiconductor lasers with delayed optoelectronic feedback is studied numerically. The effects of positive and negative delayed optoelectronic feedback in producing chaotic outputs from such lasers with nonlinear gain reduction in its optimum value range is investigated using bifurcation diagrams. The results are confirmed by calculating the Lyapunov exponents. A negative delayed optoelectronic feedback configuration is found to be more effective in inducing chaotic dynamics to such systems with nonlinear gain reduction factor in the practical value range.

## 1. INTRODUCTION

Study of chaotic dynamics of semiconductor lasers have received much attention in the past few decades due to the applicability of chaotic synchronization of such systems in the field of optical secure communication [1-8]. Semiconductor lasers are generally very stable systems when operated with only a dc bias current. Instabilities are induced in their dynamics by inclusion of additional degrees of freedom. The different methods proposed for producing chaotic outputs are giving external optical injection [9-12], giving optical feedback [13-15], direct GHz current modulation [16-23] and giving delayed optoelectronic feedback [24-28]. Positive delayed optoelectronic feedback is the conventional method of generating ultra-short pulses [29] from semiconductor lasers. The dynamics of semiconductor lasers with direct current modulation has been widely studied [30-34]. It has already been proved that the effect of mode gain reduction occurring due to nonlinear processes is suppression of chaotic dynamics [35]. A bi-directional coupling between two such lasers is also found to suppress chaotic dynamics [36]. A positive delayed optoelectronic feedback combined with strong current modulation is found to suppress chaotic dynamics and bistabily in semiconductor lasers [37, 38]. The effect of such a combination in inducing chaotic dynamics through a quasiperiodic route in quantum-well lasers also has been studied [25].

---
[1][1]corresponding author: bindum@cusat.ac.in (nee:V.Bindu)

The most preferred light source in the optical communication systems is the directly modulated semiconductor lasers with GHz modulation. Chaotic synchronization of two such lasers is a widely investigated topic because of its applicability in optical secure communication [39-42]. For InGaAsP lasers used in optical communication systems, the nonlinear gain reduction is very strong and its direct consequence on the dynamics of such lasers is the suppression of chaotic outputs. Earlier studies on dynamics of directly modulated semiconductor lasers based on the rate equations for carrier and photon population inside the laser cavity, predicted period doubling and chaos in some range of modulation frequency and modulation depth [18, 19]. These rate equations have to be modified to include a small power dependent reduction in mode gain occurring due to phenomenon such as spectral hole burning [43, 44]. The dynamic response of semiconductor laser strongly depends on the nonlinear gain and therefore it has a significant role in modeling semiconductor laser dynamics [45, 46]. The optimum value of nonlinear gain reduction factor for InGaAsP lasers is between 0.03 and 0.06 and it has been proved that this system exhibits chaotic dynamics only for nonlinear gain reduction below 0.01 [35]. This makes the investigations on the methods of producing chaotic outputs from such lasers under optimum parameter values very important for their applicability in secure communication systems.

Here we report the results of numerical investigations on the effect of a delayed optoelectronic feedback on the dynamics of such lasers. The study was undertaken to find out the possibility of obtaining chaotic outputs from InGaAsP lasers under normal values of nonlinear gain reduction factor. The nonlinear dynamics of semiconductor lasers without current modulation under positive and negative delayed optoelectronic feedback configurations has been investigated both numerically and experimentally. The results of these studies have proved that the pulsing mechanisms are not necessarily the same for these two configurations [24]. The optoelectronic feedback scheme has the advantage of ease of implementation, as it is insensitive to the optical phase of the output intensity. Therefore the effects of both a positive and negative delayed optoelectronic feedback schemes were investigated. The results reveal that in the range of normal estimates of nonlinear gain reduction factor for such lasers as suggested by Agrawal [35], only a strong negative delayed optoelectronic feedback is efficient in producing chaotic output.
The present paper is organized as follows. The model rate equations and the schematic of the optoelectronic feedback configuration are presented in §2. In §3 the results of the numerical investigation on the dynamics of directly modulated semiconductor lasers with positive and negative delayed optoelectronic feedback are presented and the concluding remarks are presented in §4.

**2. LASER MODEL**
Semiconductor lasers with direct current modulation can be represented by the following rate equations for the photon density (P), carrier density (N), and the driving current (I) [21, 35]

$$\frac{dN}{dt} = \frac{1}{\tau_e}\left\{\left(\frac{I}{I_{th}}\right) - N - \left[\frac{N-\delta}{1-\delta}\right]P\right\} \quad (1)$$

$$\frac{dP}{dt} = \frac{1}{\tau_p}\left\{\left[\frac{N-\delta}{1-\delta}\right](1-\varepsilon P)P - P + \beta N\right\} \quad (2)$$

$$I(t) = I_b + I_m Sin(2\pi f_m t) \quad (3)$$

where $\tau_e$ and $\tau_p$ are the electron and photon lifetimes, N and P are the carrier and photon densities, I is the driving current, $\delta = n_0/n_{th}$, $\varepsilon = \varepsilon_{NL}S_0$ are dimensionless parameters where $n_0$ is the carrier density required for transparency, $n_{th} = (\tau_e I_{th}/eV)$ is the threshold carrier density, $\varepsilon_{NL}$ is the factor governing the nonlinear gain reduction occurring with an increase in S, $S_0 = \Gamma(\tau_p/\tau_e)n_{th}$, $I_{th}$ is the threshold current, e is the electron charge, V is the active volume, $\Gamma$ is the confinement factor and $\beta$ is the spontaneous emission factor. $I_b = b \times I_{th}$ is the bias current where b is the bias strength, $I_m = m \times I_{th}$ is the modulation current where m is the modulation depth and $f_m$ is the modulation frequency [35]. Parameter values for which the system output will be chaotic are given in Table.1.

The schematic diagram of directly modulated semiconductor lasers with a delayed optoelectronic feedback is shown in Fig. 1. The laser diode (LD) is modulated with a sinusoidal current of GHz frequency. The light output from the laser diode is converted into electronic signal using a photo detector (PD) and then amplified to the required gain. This current is fed back to the input of the laser diode in addition to its injection current. Allowing the light to travel a certain distance through free space before reaching the PD can provide the required delay. For GHz modulation, the delay is of the order of nanoseconds that correspond to a traveling distance of the order of a few centimeters. For a positive delayed optoelectronic feedback, the feedback current is added to the input in addition to the bias and modulation currents, whereas for a negative feedback it is deducted from the total input current comprising of the bias and modulation currents. Correspondingly the equation for the input current will be modified as follows.

$$I(t) = I_b + I_m Sin(2\pi f_m t) \pm r \times (P_{t-\tau}) \quad (4)$$

where r is the feedback fraction and $\tau$ is the delay time. The equations (1), (2) & (4) modeling the dynamics of laser diode with delay feedback is solved numerically using fourth-order Runge-Kutta algorithm with parameter values for chaotic outputs as given in Table.1.

3. RESULTS AND DISCUSSION

Chaotic dynamics of directly modulated semiconductor lasers with GHz modulation has been studied in detail and the condition for occurrence of chaotic output derived analytically in [35]. It has been proved that the output dynamics of such systems will show chaotic nature for a small window of modulation depth when the nonlinear gain reduction factor is very low. Fig. 2a shows the variation in output photon density of this system without a feedback at nonlinear gain reduction factor of $\varepsilon = 0.0001$, as modulation depth is increased from zero to 0.1. The maxima of the output photon density inside each modulation cycle are plotted against the corresponding modulation depth. For $\varepsilon = 0.0001$, the output dynamics traces a period doubling route to chaotic nature at modulation depths between 0.45 and 0.6. For further increase of modulation depth, the dynamics traces a reverse period doubling route to stability. Fig. 2b shows the effect of increase in the nonlinear gain reduction factor on the output dynamics when all other control parameters are kept constant at the chaotic operating condition as in Table.1. It is clear from the bifurcation diagram in Fig. 2b that the output undergoes a reverse period doubling for an increase in the nonlinear gain reduction factor and attains a stable nature at $\varepsilon = 0.01$. These results agree with the condition for the stability of the periodic solution of system of equations (1), (2) & (3) earlier reported [35].

To study the effect of both positive and negative delayed optoelectronic feedback for systems with higher nonlinear gain reduction factor, the dynamics for different values of time delay and feedback fractions for systems with nonlinear gain reduction in the range $0.0001 \leq \varepsilon \leq 0.1$ are simulated giving special attention to the range $0.03 \leq \varepsilon \leq 0.06$. Fig. 3a and Fig. 3b show the bifurcation diagrams of the output dynamics at $\varepsilon = 0.0001$ for positive and negative feedback configurations respectively. The feedback fraction is kept at a low value of $r = 0.005$. The delay time is increased from zero to a high value of $4ns$ for both positive and negative feedback configurations. It can be seen that as the delay time increases, the output shows a rich variety of dynamics upto $\tau = 1.25ns$ which is the inverse of the modulation frequency, $f_m = 0.8GHz$, in the case of a positive delayed optoelectronic feedback. The output dynamics which is chaotic for zero feedback delay time becomes periodic for a delay time as short as 0.05ns. With further increase in the delay time the output undergoes period doubling and attains chaotic state at 0.5ns which lasts upto 0.58ns. The time series of output photon density is extracted and the Lyapunov exponents per time step of 0.01ns are calculated using the software TISEAN.7.0. Time delayed mutual information and false nearest neighbour methods are used to estimate the delay and embedding dimension for attractor reconstruction. The largest Lyapunov exponent at $\varepsilon = 0.0001$, r=0.005, and $\tau = 0.05ns$ is 0.0626. The irregular behaviour between 0.58ns and 0.65ns in the bifurcation diagram represents quasiperiodic states. Fig. 3c show the frequency spectrum at $\varepsilon = 0.0001$, r=0.005, and $\tau = 0.58ns$. Between 0.65ns and 1.25ns of delay time the output is periodic and between 1.25ns and 1.51ns the output is chaotic. The largest Lyapunov exponent at $\tau = 1.25ns$ is 0.0502. The chaotic dynamics reappears at 1.64ns after a periodic region from 1.51ns to 1.64ns and sustains upto 1.71ns. Thereafter the output dynamics is periodic upto 1.79ns. The dynamic behaviour between 1.79ns and 1.9ns is quasiperiodic. Similar chaotic and quasiperiodic regimes are repeated at higher time delays also.

In the case of negative delayed feedback configuration shown in Fig. 3b, the output dynamics remains chaotic upto 0.03ns where it changes to regular single periodic state. This state is maintained for higher values of delay times upto 1.11ns. The output dynamics is chaotic between 1.11ns and 1.28ns. The largest Lyapunov exponent at $\varepsilon$ = 0.0001, r= 0.005 $\tau$ = 1.25*ns* is 0.0448. Similar behaviour is observed for higher delay time intervals.

Keeping the feedback fraction at *r* = 0.005 and feedback delay time at 3.78*ns*, which is high above 1.25ns and where the dynamics is chaotic for both positive and negative configurations at $\varepsilon$ = 0.0001, equations (1), (2) & (4) are simulated for higher values of nonlinear gain suppression factor. The output peaks are detected from each modulation cycle and are plotted against the corresponding values of nonlinear gain reduction factor. Fig. 4a shows the dynamics of positive feedback configuration and Fig. 4b that of negative feedback configuration. It can be seen that for both configurations this feedback can induce chaotic dynamics for $\varepsilon$ < 0.02. However, at $\varepsilon$ = 0.01 where the output is stable for lasers without feedback, the negative feedback scheme could induce chaotic dynamics.

To study the effect of feedback delay time on the output dynamics at $\varepsilon$ = 0.05, the bifurcation diagrams of the output were plotted for increasing values of time delays keeping the feedback fraction at a constant low value of *r* = 0.005. Fig. 5a and Fig. 5b show the bifurcation diagrams for positive and negative configurations respectively. It can be seen that the dynamics is at a stable point for both cases for all values of delay times for *r* = 0.005. With further increase of feedback fraction to r=0.02, the output dynamics of laser with negative delayed optoelectronic feedback is found to become chaotic whereas the positive feedback configuration is not found to be effective in producing chaotic outputs. Fig. 6a and Fig. 6b show the bifurcation diagrams of output dynamics at $\varepsilon$ = 0.05 and r=0.02, for positive and negative delayed feedback configurations respectively. The delay time is increased from zero to a high value of 4ns in both the cases. For the negative feedback configuration, the output dynamics which is in a regular period one state for zero time delay switches to two frequency quasiperiodic state at 1.19ns and again becomes regular period one at 1.32ns. With further increase in delay time the output again changes to a three frequency quasiperiodic state at 2.1ns and then to chaotic state at 2.3ns. This chaotic behaviour is maintained upto a delay time of 2.51ns where the dynamics switches to a three frequency quasiperiodic state and subsequently to regular period one state at 2.6ns. The largest Lyapunov exponent at $\varepsilon$ = 0.05, r=0.02 and $\tau$ = 2.5*ns* is 0.0626. Similar periodic, quasiperiodic and chaotic behaviour are observed at higher delay time intervals. These results clearly prove that a negative delayed optoelectronic feedback is the most effective method of producing chaotic outputs from directly modulated semiconductor lasers under optimum values of nonlinear gain reduction factor.

To investigate the effect of increase in the feedback fraction on the output dynamics, the delay time was kept constant at a fairly high value of $\tau$ = 3.78*ns* where both configurations are effective in producing chaotic dynamics (at lower values of nonlinear gain reduction) and the bifurcation diagrams were plotted for increasing values of feedback fraction. Fig. 7a and Fig. 7b show the bifurcation for positive and negative feedback configurations. It is evident from these figures that a negative delayed optoelectronic feedback is more effective in inducing chaotic dynamics in the laser output at the optimum value of nonlinear gain suppression. In the case of negative delayed feedback, as the feedback fraction is increased the dynamics which is regular single periodic passes through quasiperiodic states to attain chaotic state at a feedback fraction of 0.017. This chaotic state is sustained upto a feedback fraction of 0.021, where it gets interrupted by periodic states upto 0.024. At this point the dynamics again switches to chaotic state and remains chaotic upto 0.028. From these results we can infer that the optimum minimum range of feedback fraction for chaotic dynamics in this case lies between 0.017 and 0.021. Fig. 7c and Fig. 7d show the power spectra of outputs from positive and negative delay systems at $\varepsilon = 0.05$, r=0.02 and $\tau = 3.78 ns$. It is clear from these spectra that only a negative delayed optoelectronic feedback is efficient in producing broadband chaotic output at $\varepsilon = 0.05$. The above results indicate that a negative delayed optoelectronic feedback is very effective in producing chaotic outputs from directly modulated semiconductor lasers under optimum values of nonlinear gain reduction factor.

**4. CONCLUSION**

A delayed optoelectronic feedback scheme is employed to study the possibilities of obtaining chaotic output from a directly modulated semiconductor laser under GHz modulation and the reported experimental value of nonlinear gain suppression factor for InGaAsP. At low values of nonlinear gain suppression factor a positive delayed feedback could produce a rich variety of dynamics in the output of this system. But, as the nonlinear gain reduction increases, the output dynamics of the system with positive feedback configuration becomes stable. However, the dynamics of the system with negative feedback is chaotic in this range of values also. The results indicate that for the optimum value range of nonlinear gain reduction factor, a negative delayed feedback configuration is effective in producing chaotic outputs. At higher values of nonlinear gain reduction factor, the negative feedback could still induce chaotic dynamics whereas the positive feedback is not found to be effective in producing chaotic dynamics. The results of the present study indicate that there is a difference in the dynamical mechanism between positive and negative delayed feedback configurations in semiconductor lasers with direct current modulation. For an explanation of the correct mechanisms effective in these dynamics and establishing the stability criteria for these two cases, a detailed analytical treatment and stability analysis of the delayed system rate equation are necessary. This work is in progress.

**ACKNOWLEDGEMENTS**

One of the authors BMK would like to acknowledge with thanks the financial support from the Department of Science and Technology (Govt. of India), through Fast Track Scheme for Young Scientists, No.SR/FTP/PS-14/2004, Prof. Jacob Philip, Director, STIC for technical support, Prof. R. Pratap, International School of Photonics, Cochin University of Science and Technology and Dr. S. Sivaprakasam, University of Pondichery, for fruitful discussions. MPJ would like to acknowledge the financial support from Council of Scientific and Industrial Research, India through Senior Research Fellowship.

**Table I: Parameter Values Used for Numerical Simulation**

| Parameter | Value |
|---|---|
| $\tau_e$ | 3ns |
| $\tau_p$ | 6ps |
| $\varepsilon$ | $1 \times 10^{-4}$ |
| $\delta$ | $692 \times 10^{-3}$ |
| $\beta$ | $5 \times 10^{-5}$ |
| $f_m$ | 0.8GHz |
| $I_{th}$ | 26mA |
| m | 0.55 |
| b | 1.5 |

**FIGURE CAPTIONS**

1. Fig.1: Schematic of directly modulated semiconductor laser with delayed optoelectronic feedback. The injection current to the laser diode (LD) is modulated by a GHz current. Light output from LD is delayed by appropriate time and is converted into electronic signal using a photodiode (PD). This current is fed back to the input of LD in addition to its injection current.
2. Fig.2a: Bifurcation diagram of output (maxima of normalized photon density, $P_{max}$) of modulated semiconductor laser for increasing values of modulation depth 'm'. The nonlinear gain reduction factor is kept at a low value of $\varepsilon = 0.0001$.
3. Fig.2b: Bifurcation diagram of output $P_{max}$, of modulated semiconductor laser without feedback for increasing values of nonlinear gain reduction factor $\varepsilon$. The modulation depth is kept at m=0.55.
4. Fig.3a: Bifurcation diagram of output $P_{max}$, of modulated semiconductor laser with positive delayed optoelectronic feedback of strength r=0.005 and nonlinear gain reduction factor of $\varepsilon = 0.0001$, for increasing values of delay time $\tau$.
5. Fig.3b: Bifurcation diagram of output $P_{max}$, of modulated semiconductor laser with negative delayed optoelectronic feedback of strength r=0.005 and nonlinear gain reduction factor of $\varepsilon = 0.0001$, for increasing values of delay time $\tau$.

6. Fig.3c: Power spectrum of output of modulated semiconductor laser at nonlinear gain reduction factor value of $\varepsilon = 0.0001$ with positive delayed optoelectronic feedback of strength r=0.005 and delay time of $\tau = 0.58ns$.
7. Fig.4a: Bifurcation diagram of output $P_{max}$, of modulated semiconductor laser with positive delayed optoelectronic feedback of strength r=0.005 and delay time $\tau = 3.78ns$, for increasing values of nonlinear gain reduction factor $\varepsilon$.
8. Fig.4b: Bifurcation diagram of output $P_{max}$, of modulated semiconductor laser with negative delayed optoelectronic feedback of strength r=0.005 and delay time $\tau = 3.78ns$, for increasing values of nonlinear gain reduction factor $\varepsilon$.
9. Fig.5a: Bifurcation diagram of output $P_{max}$, of modulated semiconductor laser at nonlinear gain reduction factor value of $\varepsilon = 0.05$, with positive delayed optoelectronic feedback of strength r=0.005, for increasing values of delay time $\tau$.
10. Fig.5b: Bifurcation diagram of output $P_{max}$, of modulated semiconductor laser at nonlinear gain reduction factor value of $\varepsilon = 0.05$ with negative delayed optoelectronic feedback of strength r=0.005, for increasing values of delay time $\tau$.
11. Fig.6a: Bifurcation diagram of output $P_{max}$, of modulated semiconductor laser at nonlinear gain reduction factor value of $\varepsilon = 0.05$ with positive delayed optoelectronic feedback of strength r=0.02, for increasing values of delay time $\tau$.
12. Fig.6b: Bifurcation diagram of output $P_{max}$, of modulated semiconductor laser at nonlinear gain reduction factor value of $\varepsilon = 0.05$ with negative delayed optoelectronic feedback of strength r=0.02, for increasing values of delay time $\tau$.
13. Fig.7a: Bifurcation diagram of output $P_{max}$, of modulated semiconductor laser at nonlinear gain reduction factor value of $\varepsilon = 0.05$ with positive delayed optoelectronic feedback, for increasing values of feedback strength r. The delay time is $\tau = 3.78ns$.
14. Fig.7b: Bifurcation diagram of output $P_{max}$, of modulated semiconductor laser at nonlinear gain reduction factor value of $\varepsilon = 0.05$ with negative delayed optoelectronic feedback, for increasing values of feedback strength r. The delay time is $\tau = 3.78ns$.
15. Fig.7.c: Power spectrum of output of modulated semiconductor laser at nonlinear gain reduction factor value of $\varepsilon = 0.05$ with positive delayed optoelectronic feedback of strength r=0.02 and delay time of $\tau = 3.78ns$.
16. Fig.7.d: Power spectrum of output of modulated semiconductor laser at nonlinear gain reduction factor value of $\varepsilon = 0.05$ with negative delayed optoelectronic feedback of strength r=0.02 and delay time of $\tau = 3.78ns$.

Fig.1

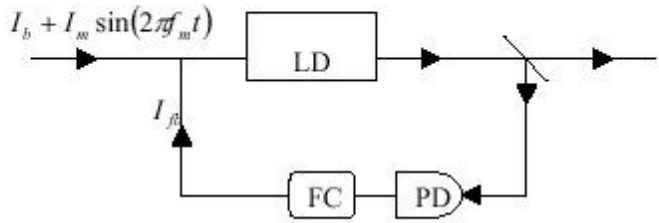

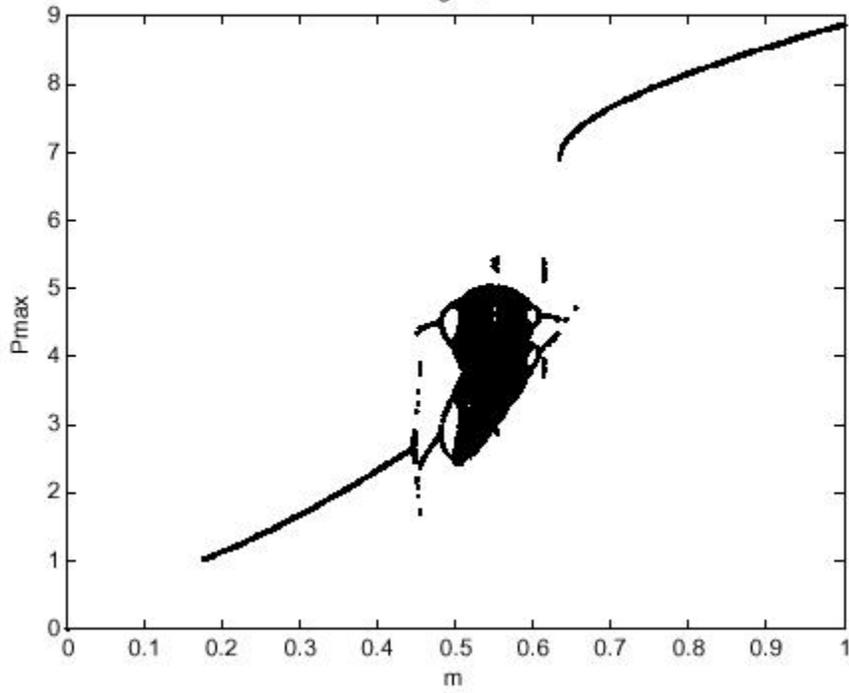

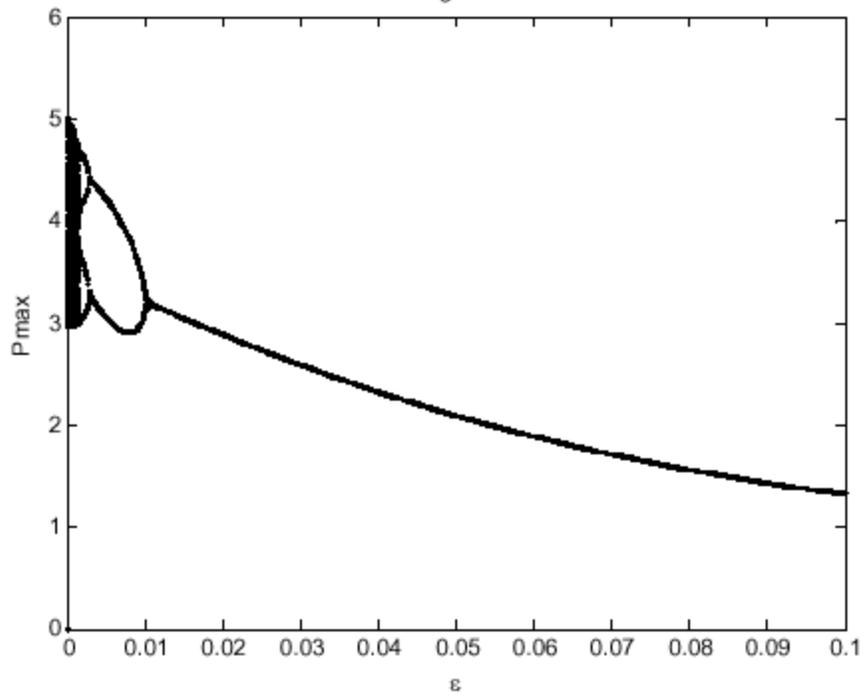

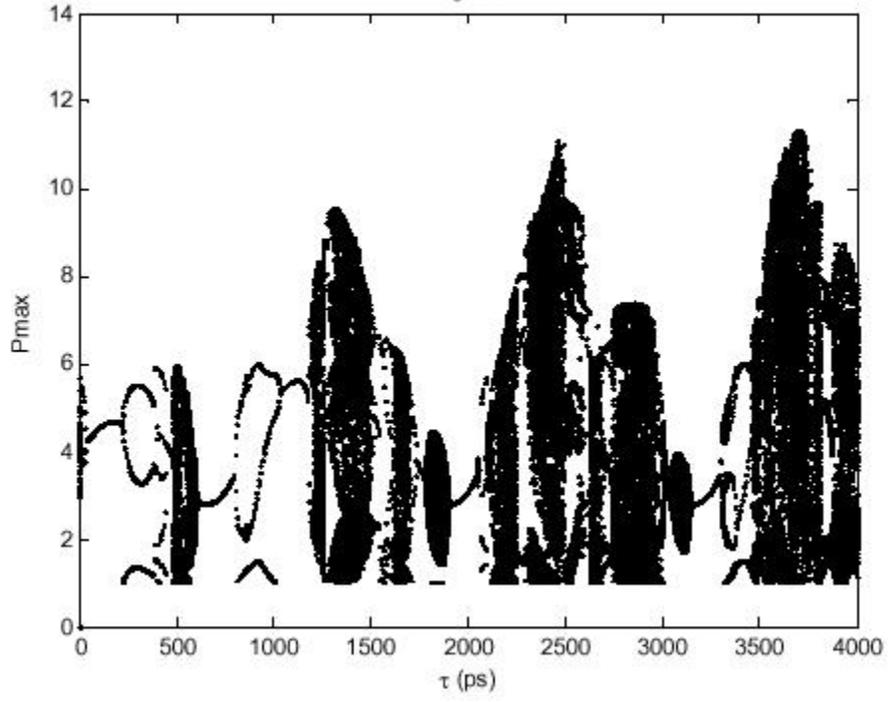

Fig. 3a

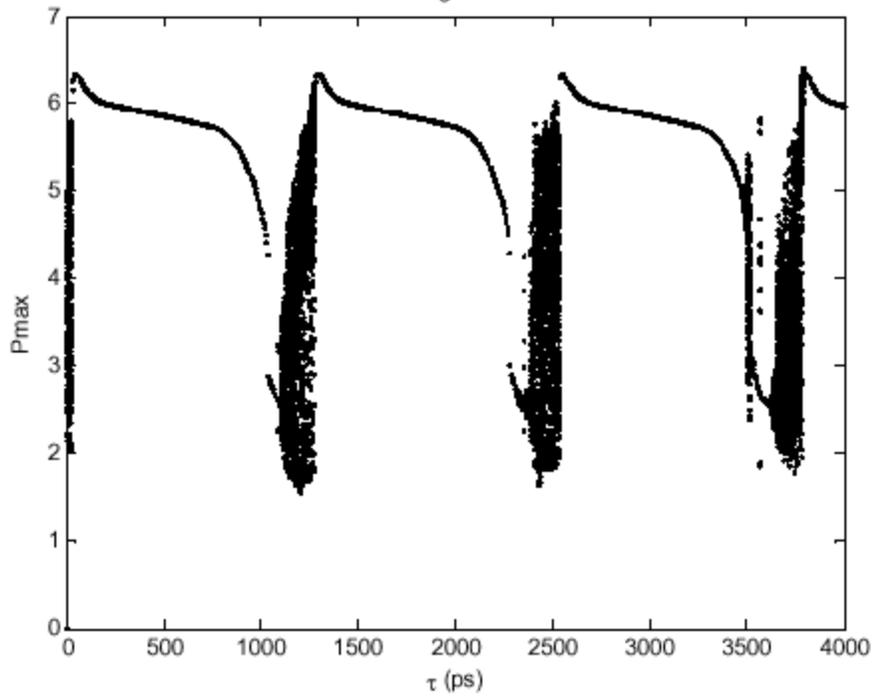

Fig. 3b

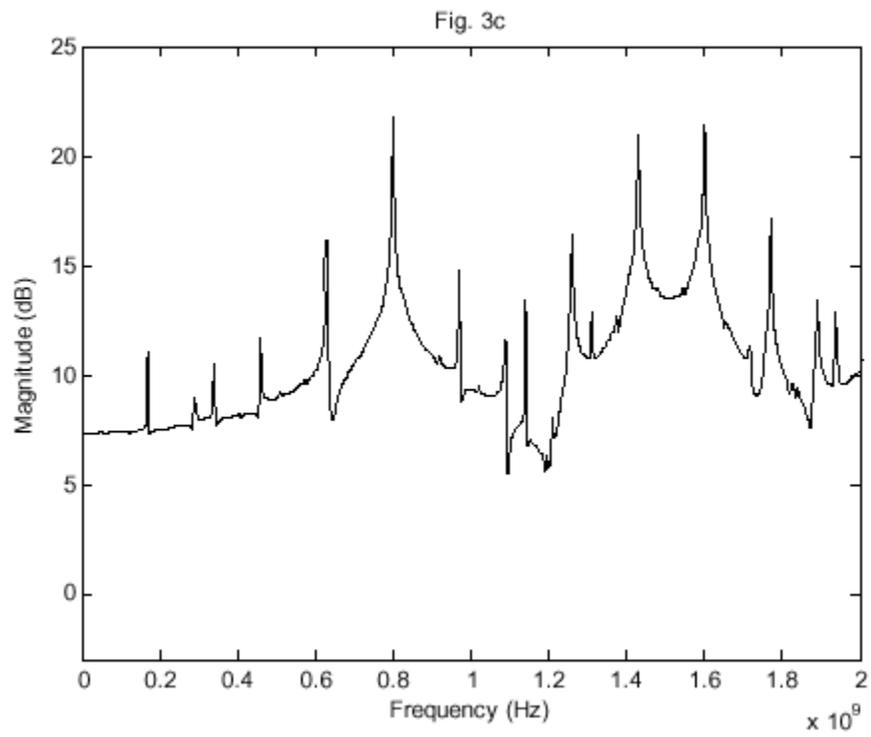

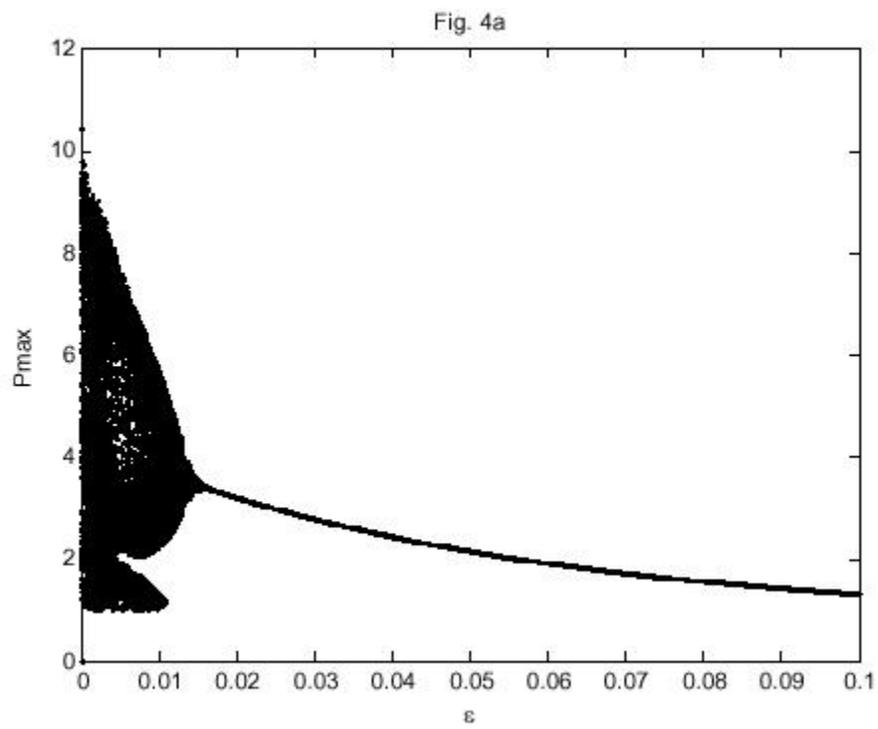

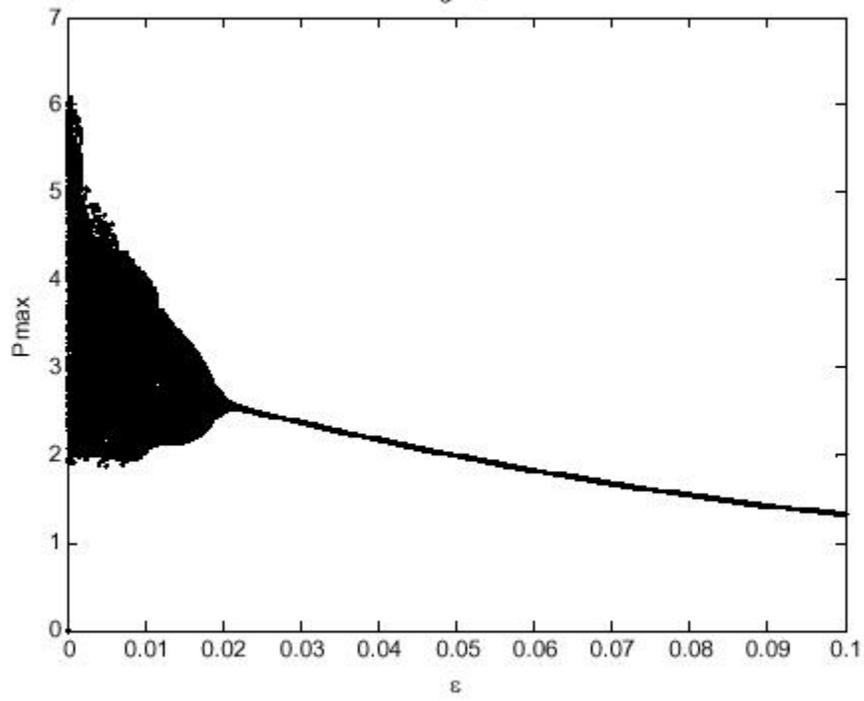

Fig. 4b

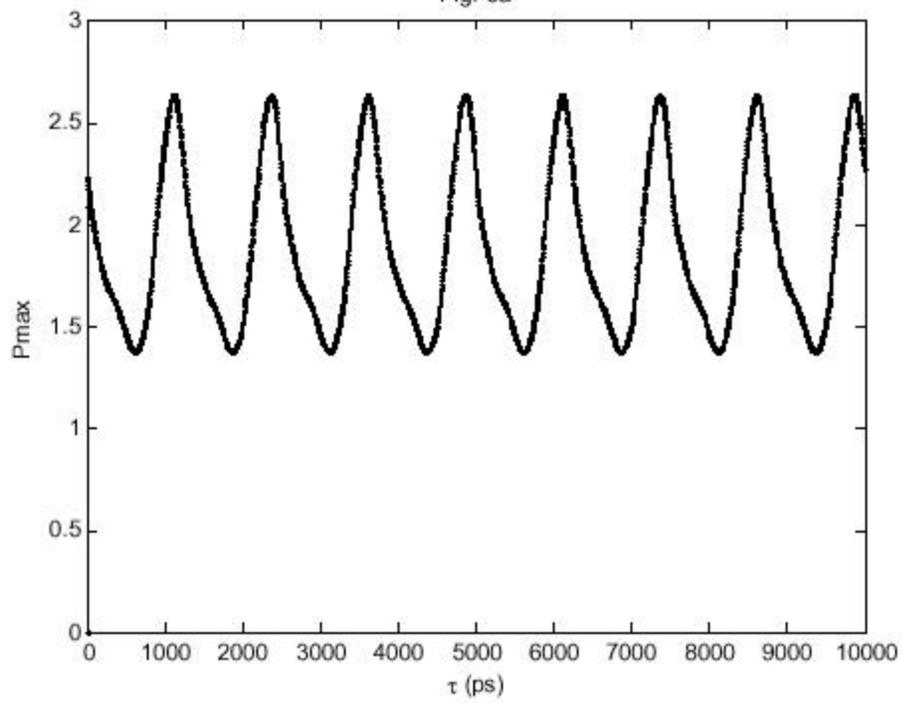

Fig. 5a

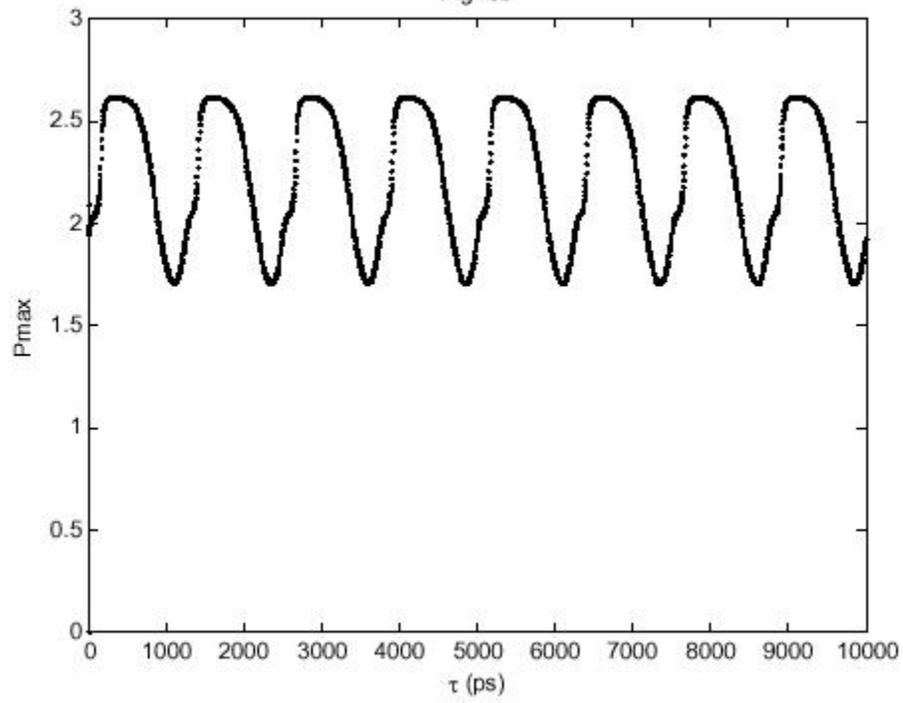

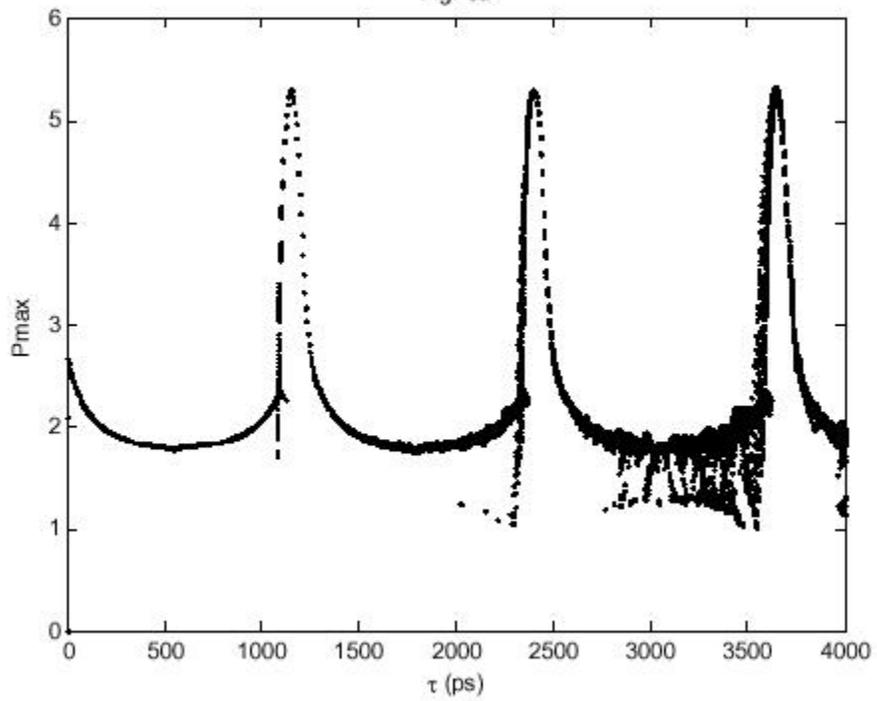

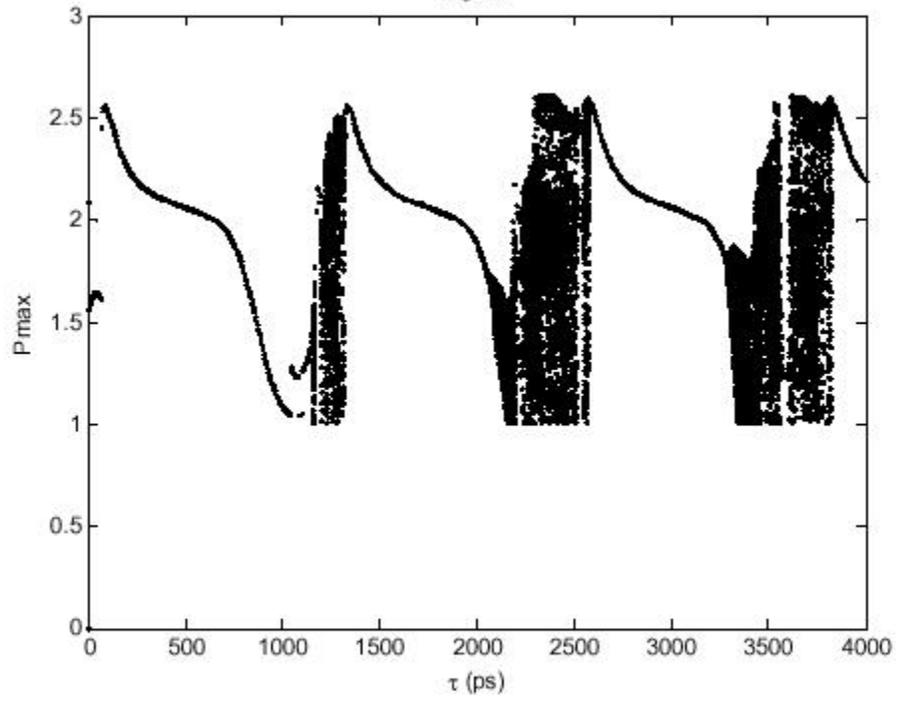

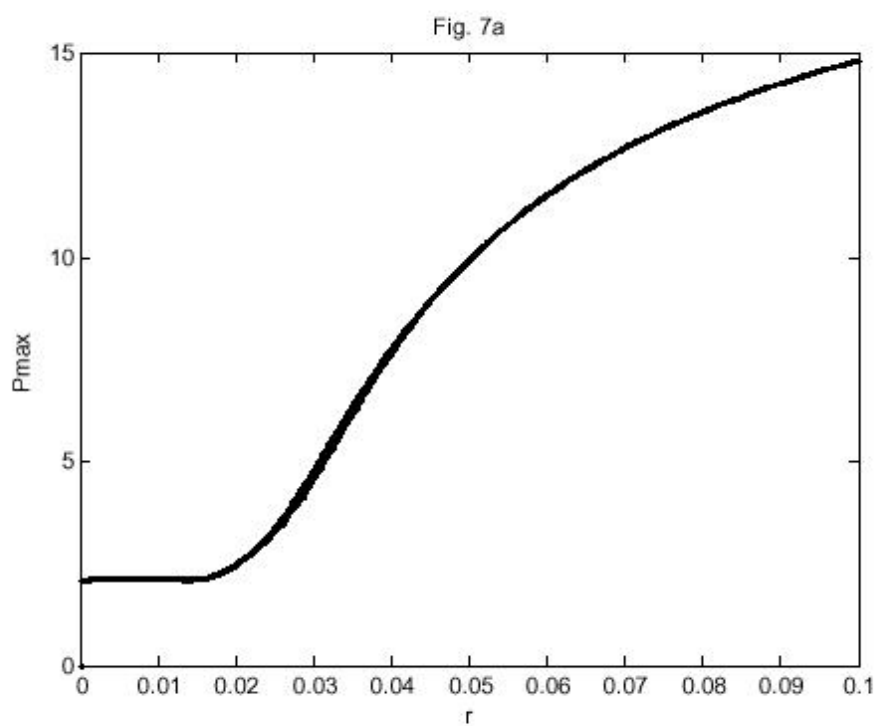

Fig. 7a

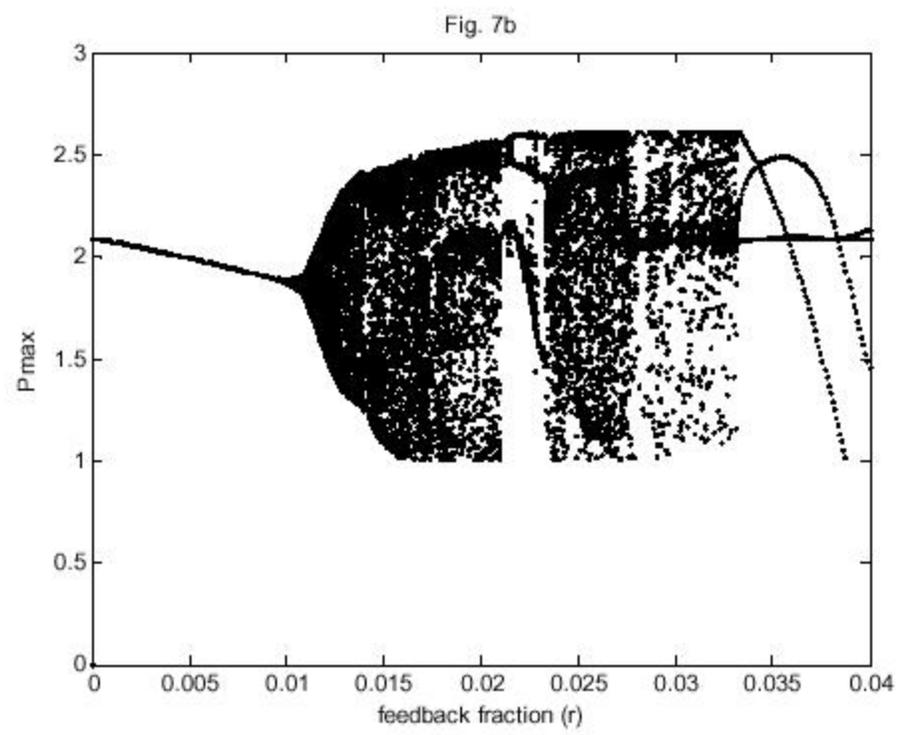

Fig. 7b

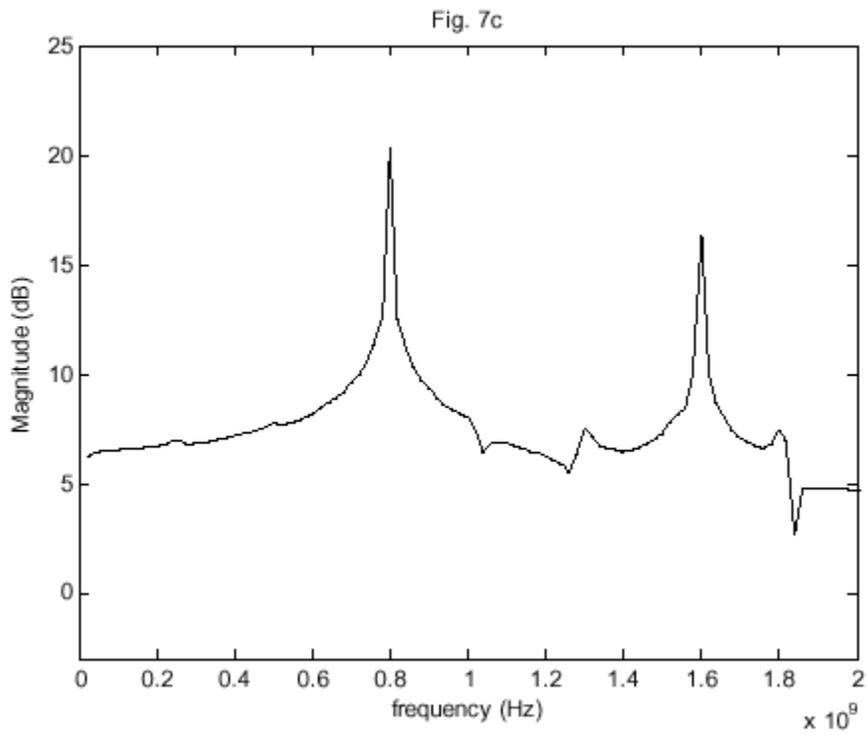

Fig. 7c

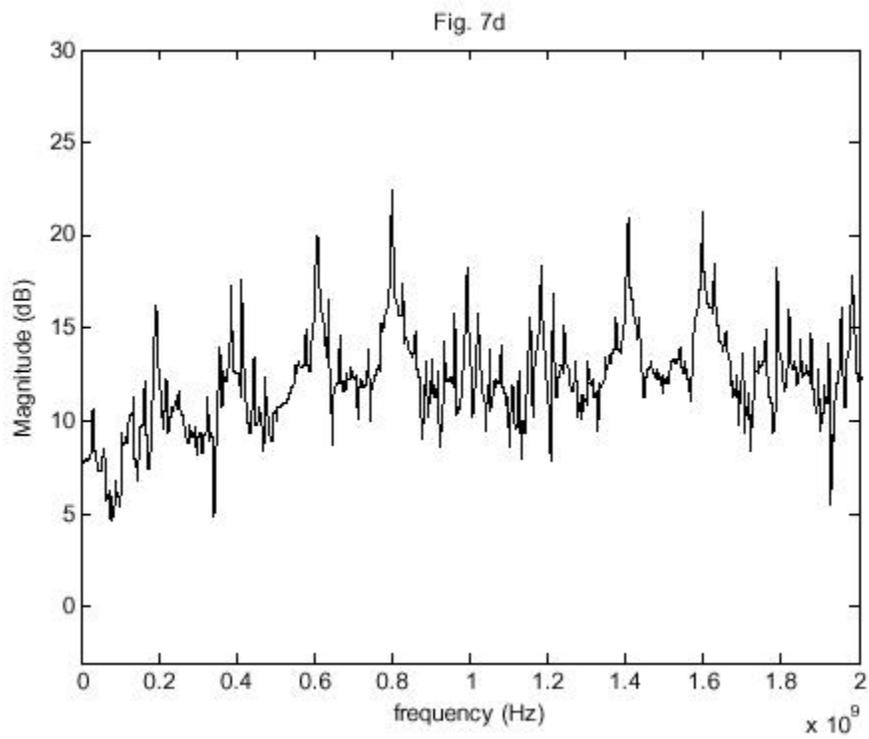

Fig. 7d